\documentclass[useAMS,usenatbib]{mn2e}

\usepackage[T1]{fontenc}
\usepackage[shortlabels]{enumitem}
\usepackage{pslatex}
\usepackage{amsmath}
\usepackage{amsfonts}
\usepackage{color}
\usepackage{url}
\usepackage{graphicx}
\usepackage{subfigure}
\usepackage{amssymb}
\usepackage{textcomp}
\usepackage{soul}
\usepackage{bm}
\usepackage{booktabs}
\usepackage{array}
\usepackage[draft]{hyperref}
 \graphicspath{{images/}}

{\newif\ifnotend
\notendtrue
\def\veclist{ABCDEFGHIJKLMNOPQRSTUVWXYZabcdefghijklmnopqrstuvwxyz.}
\def\top#1#2.{#1}
\def\tail#1#2.{#2.}
\loop\expandafter\xdef\csname v\expandafter\top\veclist\endcsname%
{{\noexpand\bf\expandafter\top\veclist}}
\edef\veclist{\expandafter\tail\veclist}
\if\veclist.\notendfalse\fi\ifnotend\repeat}

\def\pa{\partial}

\mathchardef\mhyphen="2D

\title[The effect of rotation on the formation of HVCs]{The effect of rotation on the thermal instability of stratified galactic atmospheres - II. The formation of High Velocity Clouds}

\author[Sormani \& Sobacchi]{Mattia C. Sormani$^{1}$\thanks{Contact email: mattia.sormani@uni-heidelberg.de} and Emanuele Sobacchi$^{2,3}$\thanks{Contact email: sobacchi@post.bgu.ac.il}\\
$^1$Universit\"{a}t Heidelberg, Zentrum f\"{u}r Astronomie, Institut f\"{u}r theoretische Astrophysik, Albert-Ueberle-Str. 2, 69120 Heidelberg, Germany \\
$^2$Physics Department, Ben-Gurion University, P.O.B. 653, Beer-Sheva 84105, Israel \\
$^3$Department of Natural Sciences, The Open University of Israel, 1 University Road, P.O.B. 808, Raanana 4353701, Israel
}

\begin{document}

\date{}

\def\p{\partial}
\def\Omegap{\Omega_{\rm p}}

\newcommand{\di}{\mathrm{d}}
\newcommand{\bfx}{\mathbf{x}}
\newcommand{\bfe}{\mathbf{e}}
\newcommand{\bfxi}{\bm{\xi}}
\newcommand{\bfk}{\mathbf{k}}
\newcommand{\vlos}{\mathrm{v}_{\rm los}}
\newcommand{\Tspin}{T_{\rm s}}
\newcommand{\Tb}{T_{\rm b}}
\newcommand{\degree}{\ensuremath{^\circ}}
\newcommand{\Th}{T_{\rm h}}
\newcommand{\Tc}{T_{\rm c}}
\newcommand{\bfr}{\mathbf{r}}
\newcommand{\bfv}{\mathbf{v}}
\newcommand{\bfu}{\mathbf{u}}
\newcommand{\bfg}{\mathbf{g}}
\newcommand{\pc}{\,{\rm pc}}
\newcommand{\kpc}{\,{\rm kpc}}
\newcommand{\Myr}{\,{\rm Myr}}
\newcommand{\Gyr}{\,{\rm Gyr}}
\newcommand{\kms}{\,{\rm km\, s^{-1}}}
\newcommand{\de}[2]{\frac{\partial #1}{\partial {#2}}}
\newcommand{\cs}{c_{\rm s}}
\newcommand{\rb}{r_{\rm b}}
\newcommand{\rqu}{r_{\rm q}}
\newcommand{\nuP}{\nu_{\rm P}}
\newcommand{\thetaobs}{\theta_{\rm obs}}
\newcommand{\hatn}{\hat{\textbf{n}}}
\newcommand{\hatt}{\hat{\textbf{t}}}
\newcommand{\hatx}{\hat{\textbf{x}}}
\newcommand{\haty}{\hat{\textbf{y}}}
\newcommand{\hatz}{\hat{\textbf{z}}}
\newcommand{\hatX}{\hat{\textbf{X}}}
\newcommand{\hatY}{\hat{\textbf{Y}}}
\newcommand{\hatZ}{\hat{\textbf{Z}}}
\newcommand{\hatN}{\hat{\textbf{N}}}
\newcommand{\hater}{\hat{\mathbf{e}}_r}
\newcommand{\hateR}{\hat{\mathbf{e}}_R}
\newcommand{\hatephi}{\hat{\mathbf{e}}_\phi}
\newcommand{\hatez}{\hat{\mathbf{e}}_z}
\newcommand{\hateP}{\hat{\mathbf{e}}_P}
\newcommand{\hatePhi}{\hat{\mathbf{e}}_\Phi}
\newcommand{\hatetheta}{\hat{\mathbf{e}}_\theta}
\newcommand{\hatemu}{\hat{\mathbf{e}}_\mu}
\newcommand{\hatenu}{\hat{\mathbf{e}}_\nu}
\newcommand{\hatePL}{\hat{\mathbf{e}}_{P\Lambda}}
\newcommand{\nablaPL}{\nabla_{P\Lambda}}

\maketitle

\begin{abstract}
Whether High Velocity Clouds (HVCs) can form by condensation of the hot ($T \sim 10^6 \, {\rm K}$) Galactic corona as a consequence of thermal instabilities has been controversial. Here we re-examine this problem and we suggest that rotation of the corona might be a missing key ingredient. We do this by studying the evolution of the models of rotating galactic coronae presented in \citet{Sormani+2018c} under the presence of cooling and thermal conduction. We combine a linear stability analysis with the results of local and global hydrodynamical simulations. We find that condensations are likely to occur in regions where the corona has substantial rotational support. Under reasonably general assumptions on the rotation profile of the corona, the locations where condensations are expected are in remarkable agreement with the observed location of the major non-magellanic HVCs complexes in our Galaxy (namely, at distances $\lesssim 15 \kpc$ from the Sun and within $30\degree$ from the disc plane). We conclude that HVCs can form by thermal instabilities provided that (i) the corona is rotating substantially in the inner ($R \lesssim 50\kpc$) parts, as suggested by current observational data and predicted by cosmological simulations of galaxy formation; (ii) close to the disc the corona is well-represented by a nearly-equilibrium stratified rotating structure (as opposed to a fast cooling flow). Our results also suggest that a better understanding of the disc-halo interface, including supernova feedback, is critical to understand the origin of HVCs.
\end{abstract}


\section{Introduction}

It is now well established that the Milky Way (MW) is surrounded by a hot gaseous atmosphere, which is known as the Galactic corona \citep[e.g.][and references therein]{Putman+2012,Tumlinson+2017,Bregman+2018}. The Galactic corona has a typical temperature of $T \simeq 2 \times 10^6 \rm \, K$ and is presumed to extend up to the virial radius of the MW dark matter halo, $r_{\rm vir} \simeq 250 \kpc$, although due to the difficulty of observations most direct detections only probe the innermost $r \lesssim 50\kpc$  \citep[e.g.][]{Yoshino2009,Gupta+2012,MillerBregman2013,MillerBregman2015}.

High Velocity Clouds (HVC) are a number of cold discrete gas clouds which are found throughout the inner portions of the Galactic corona \citep[e.g.][]{Putman+2012,Westmeier2018}.\footnote{In this paper we consider HVC to be a population distinct from that of the Intermediate Velocity Clouds (IVCs), which are found closer to the Galactic disc and constitute the so-called \emph{extraplanar} H{\sc I} \emph{gas} \citep[e.g.][for a recent review]{Fraternali2017}. Most of the gas that constitutes IVCs has been probably pushed off the Galactic disc by stellar feedback associated with star formation \citep[e.g.][]{FraternaliBinney2008,Marasco+2011,Marasco+2012}, as opposed to HVCs that are probably of external origin as implied by their low metallicities (see below).} They are primarily detected through the HI 21cm line and are characterised by radial velocities that are inconsistent with the circular rotation of the Galactic disc. Analogous populations of cold gas clouds are observed in M31 and other nearby spiral galaxies \citep[e.g.][]{Thilker+2004,Westmeier+2008,Lockman2017}.

The HVCs have the following properties: 
\begin{enumerate}
\item They are much colder than the surrounding hot medium which envelopes them;
\item They have pressures which are roughly consistent with pressure equilibrium with the surrounding hot medium \citep[e.g.][]{Stanimirovic+2002,Fox+2005};
\item They typically have low metallicities, in the range of $0.1 \mhyphen 0.3 \,  Z_\odot$ \citep[e.g.][]{VanWoerdenWakker2004};
\item Current observational constraint place most of the HVCs in proximity of the Galactic disc, at heights $z \lesssim 10 \kpc$, distances from the Sun between $2 < d <15 \kpc$, and within $30\degree$ from the disc plane as viewed from the Galactic centre \citep[e.g.][]{Putman+2012}. The exception is the Magellanic stream, which is clearly associated with the Magellanic clouds and is found at distance of $\sim 55\kpc$;\footnote{It should be noted that while all non-Magellanic HVCs are indeed found in proximity of the Galactic plane, a significant amount of low velocity cold gas in the CGM (even close to the viral radius) may be missed by current observations \citep[e.g.][]{Zheng2015}.}
\item They seem to be infalling with typical velocities of the order of $v\sim \text{tens of }\kms$ \citep{Fox+2004,Putman+2012}.
\end{enumerate}

The origin of HVCs is not fully understood. Several hypotheses have been put forward including:\footnote{Interestingly, most of the hypotheses still discussed today were already considered by \cite{Oort1966} in the very early days (see his section 2).}
\begin{enumerate}[(a)]
\item They are made of gas ejected from the Galactic disc by supernova explosions through the galactic fountain mechanism \citep[e.g.][]{Bregman1980}. This possibility is currently disfavoured by the low metallicities measured in the main HVC complexes (see item iii above), which strongly point to an extragalactic origin;
\item They are gas that has been stripped from satellites, and/or from gas that is coming onto the Milky Way along cold cosmological filamentary streams \citep[e.g.][]{Mayer+2006, Olano2008, Fernandez+2012};
\item They are formed through condensation of the Galactic corona via thermal instabilities \citep[e.g.][]{MallerBullock2004};
\item They are formed through condensation of the Galactic corona, which however do not occur spontaneously but need to be `seeded' by the blowout of a particularly powerful superbubble that reaches high altitudes above the disc \citep{Fraternali+2015}.
\end{enumerate} 

The interesting possibility (c) was supported by early cosmological simulations \citep{Kaufmann+2006,Peek+2008}. This possibility was examined critically by \citet{Binney+2009} which, by performing a linear stability analysis of a non-rotating, stratified, pressure-supported atmosphere, found that thermal instability is strongly suppressed by a combination of buoyancy and thermal conduction and therefore concluded that it is an unlikely mechanism for the formation of the HVCs. In support of this it has been discovered that the cold clumps seen in the earlier simulations were numerical artefacts \citep{Hobbs+2013}. However, newer state of the art cosmological simulations at high resolution show the widespread presence of neutral hydrogen cloudlets in the circumgalactic medium \citep[e.g.][]{Vandevoort+2019}.

One of the critical assumptions in the work by \citet{Binney+2009}, as mentioned in their conclusions, is that the corona does not rotate. However, there is robust evidence that the Galactic corona rotates. For example, by measuring the Doppler shifts of the O{\sc VII} absorption lines, \citet{HodgesKluck+2016} found that the innermost parts of the Galactic corona rotate at $\sim180\kms$ and contain an amount of angular momentum comparable to that in the stellar disc. \citet{Marinacci+2011} used wind-tunnel simulations of cold clouds travelling through a hot medium to show that if the inner corona of a Milky Way-like galaxy did not rotate, then it would be rapidly accelerated by interaction with the extra-planar HI gas. They concluded that the corona is expected to rotate and to lag in the inner regions by $\sim 80 \kms$ with respect to the cold disc. By analysing a sample of simulated galaxies from the Eagle cosmological simulations, \citet{Oppenheimer2018} concluded that galactic coronae have significant rotational support in the inner parts (at radii $\ll$ than the virial radius).

\citet{Nipoti2010} extended the work of \citet{Binney+2009} and investigated whether rotation might change the stability properties of the Galactic corona. Although this author formally derived the relevant dispersion relations, his analysis was inconclusive because of a lack of explicit models of rotating Galactic coronae to which he could apply his analysis to. Moreover, it is now clear that the results of the linear theory are of limited applicability to quantify the condensation of non-linear structures \citep[see for example][]{McCourt+2012, Sharma+2012}. 

This motivates a re-examination of the question of whether condensation can form via thermal instabilities from a rotating Galactic corona. In this paper we use the models of \citet{Sormani+2018c} to study this question. In the absence of cooling and thermal conduction these models are stable equilibria which rotate in the inner parts. We study how they evolve when cooling and thermal conduction are added to the equations of motion. This study follows a companion paper \citep[][hereafter Paper {\sc I}]{SobacchiSormani2019a}, in which we investigated the effects of rotation on the thermal instability using an idealised plane-parallel setup, and we found that rotation enhances condensation by increasing the threshold value of the ratio between the cooling and the dynamical time below which thermal condensations can be formed.

The paper is structured as follows. In Sect. \ref{sec:physicalmodel} we describe our physical model by writing down the equations that govern the evolution of our system and the important timescales of the problem. In Sect. \ref{sec:linear} we apply the linear stability analysis of \citet{Nipoti2010} and the results of Paper {\sc I} to our models. To confirm our findings and study the non-linear behaviour, in Sect. \ref{sec:nonlinear} we perform 2.5D axisymmetric simulations which include cooling and thermal conduction using the models of \citet{Sormani+2018c} as initial conditions. We discuss our results in Sect. \ref{sec:discussion}, and we sum up in Sect. \ref{sec:conclusions}.

\section{Physical model} \label{sec:physicalmodel}

The Galactic corona is expected to be approximately in equilibrium, stratified in the gravitational potential of the Galaxy, and rotating in the inner parts (see introduction). The question is whether condensations due to the thermal instability can form in such a system. The question is \emph{not} whether the system is formally unstable according to a linear analysis, i.e. whether there exist at least one unstable mode that grows exponentially in time. Linear perturbation analysis shows that the system is always formally unstable but it has been previously argued that, despite the presence of unstable modes, the system will not develop condensations in the non-rotating case because buoyancy strongly suppresses their formation during the non-linear evolution of the system \citep{Malagoli+1987,Binney+2009}. Numerical simulations have later confirmed that indeed buoyancy can suppress the production of condensations and that non linear effects play a crucial role in the development of the instability \citep[e.g.][]{McCourt+2012}. \citet{Nipoti2010} suggested that conclusions similar to those of \cite{Binney+2009} (i.e. the corona should not develop condensations) also hold in the rotating case. However (i) the presence of additional unstable modes makes the final result more uncertain; (ii) we have found in Paper {\sc I} that condensations are more likely to form in the rotating case; (iii) the presence of rotation changes the global structure of the corona, which was not taken into account in the discussion of \cite{Nipoti2010}; (iv) Nipoti's analysis was limited to the linear regime, but we have demonstrated in Paper {\sc I} that non-linear effects play a key role also in the rotating case.

It is therefore important to (i) revisit the analysis of whether condensation can form in the Galactic corona using realistic rotating models and (ii) follow the evolution of the system in the non-linear regime. We address these items in Sect. \ref{sec:linear} and \ref{sec:nonlinear}. In the remainder of this section, we lay out the fundamental equations and the important timescales that govern the evolution of the system.

\subsection{Fundamental equations}
\label{sec:equations}

We assume that the system is governed by the equations of fluid dynamics:
\begin{align}
& \frac{\pa\rho}{\pa t} + \nabla\cdot\left(\rho{\bf v}\right)=0;,  												\label{eq:continuity}				\\ 
& \left[\frac{\pa}{\pa t} + \left({\bf v}\cdot\nabla\right)\right]{\bf v} = -\frac{\nabla P}{\rho} - \nabla\Phi\;, 					\label{eq:euler}					\\
& \frac{P}{\gamma -1}\left[\frac{\pa}{\pa t} + \left({\bf v}\cdot\nabla\right)\right]\sigma = 
\nabla\cdot\left(f \kappa_{\rm S} T^{5/2}\nabla T\right) -  \left(\frac{\rho}{\mu m_{\rm p}}\right)^2\Lambda\left(T\right)\;,		\label{eq:entropy_cons} 		
\end{align}
where $\rho$ is the density, ${\bf v}$ is the velocity, $P$ is the pressure, $\Phi$ is the external gravitational potential, $T$ is the temperature and $\sigma~=~\log\left(P \rho^{-\gamma}\right)$ is the entropy. The energy equation has been modified with respect to the ideal case by the addition of cooling and thermal conduction to the right hand side of Eq. \eqref{eq:entropy_cons}, where $\Lambda$ is the cooling function, $\kappa_{\rm S}= (5/2) k_{\rm b}^{7/2} / (\pi m_{\rm e}^{1/2} {\rm e}^4) \simeq 1.53\times 10^{-5}\;{\rm erg}\;{\rm s}^{-1}\;{\rm cm}^{-1}\;{\rm K}^{-7/2}$ is the coefficient of thermal conductivity \citep{Spitzer1962}, $m_{\rm e}$ is the electron mass, ${\rm e}$ is the electron charge, $k_{\rm b}$ the Boltzmann constant, and $f\leq 1$ is the fraction by which thermal conduction is suppressed by a tangled magnetic field \citep[e.g.][]{BinneyCowie1981}.\footnote{Of course, the suppression factor $f$ is a crude effective prescription, while a self-consistent treatment of the problem should include the effect of anisotropic thermal conduction. However, this requires to include magnetic fields in the analysis, which lies beyond the scope of the paper.} We take a fiducial $f = 0.01$ as indicated by studies of galaxy clusters \citep[e.g.][]{NipotiBinney2004} and to provide a direct comparison with the work by \cite{Binney+2009} which adopted the same value. In this paper we adopt an adiabatic index $\gamma=5/3$, the value for monoatomic ideal gases, and an ideal equation of state $P= n kT$, where $n = \rho/ (\mu m_{\rm p})$ is the number density, $\mu=0.58$ is the mean molecular weight and $m_{\rm p}$ is the proton mass. 

Throughout this paper, we adopt the collisional ionisation equilibrium cooling function from \cite{sd93} for ${\rm [Fe/H]}=-0.5$, which accounts for the sub-solar metallicities found in the HVCs and which are believed to characterise the Galactic corona \citep[e.g.][]{VanWoerdenWakker2004,Bregman+2018}.

We use standard cylindrical coordinates $\left(R,\phi,z\right)$, and we use $r$ to denote the spherical radius.

\subsection{Important timescales}

The evolution of the corona is determined by the interplay of the following timescales (see also Paper {\sc I}). The first is the dynamical timescale:
\begin{equation}
\label{eq:tdyn}
t_{\rm dyn}= \sqrt{2}\frac{c_{\rm s}}{|\bfg_{\rm eff}|}
\end{equation}
where $\cs = \sqrt{P/\rho}$ is the sound speed (apart from an unimportant factor $\gamma=5/3$) and $\bfg_{\rm eff} = - \nabla \Phi + \Omega^2 R \; \hateR$ is the effective gravity. In a nearly-equilibrium corona we have $\bfg_{\rm eff} \simeq \nabla{P}/\rho \sim P/(r \rho) \sim c_{\rm s}^2/r$ so $t_{\rm dyn} \sim \sqrt{2r / |\bfg_{\rm eff}|}$. Therefore, in the case of a non-rotating corona, $t_{\rm dyn}$ is essentially the free-fall time. In the case of a rotating corona, $t_{\rm dyn}$  it is a useful generalisation of the free-fall time that is based only on local quantities and therefore allows for a direct comparison with the local analysis of Paper {\sc I}.

The second is the rotational timescale
\begin{equation}
\label{eq:trot}
t_{\rm rot}=\frac{2\pi}{\Omega}\;,
\end{equation}
where $\Omega = v_{\rm \phi}/R$ is the local angular velocity. The third is the cooling timescale
\begin{equation}
\label{eq:tcool}
t_{\rm cool}= \frac{ 3 k_{\rm b} T }{ 2 n \Lambda(T)},
\end{equation}
where $k_{\rm b}$ is the Boltzmann constant. This is ratio between the internal energy per unit volume, $(3/2) n k_{\rm b} T$, and the amount of energy lost due to cooling per unit time and volume, $n^2 \Lambda(T)$. It therefore represents the time a blob of gas would take to emit all its energy if the cooling continued at the current rate.

Finally, we have the thermal conduction timescale:
\begin{equation}
\label{eq:tcond}
t_{\rm cond}= \frac{P L^2}{f \kappa_{\rm S} T^{7/2}} \;.
\end{equation}
This is the time it takes to thermal conduction to be effective over a region of size $L$.

\section{Thermal instability in rotating coronae} \label{sec:linear}

\subsection{Linear analysis}

The linear evolution of axisymmetric perturbations of a rotating atmosphere described by equations \eqref{eq:continuity}-\eqref{eq:entropy_cons} has been studied by \citet{Nipoti2010}. Here we follow his treatment and refer to his paper for a detailed derivation of the dispersion relation.

The unperturbed corona is assumed to be axisymmetric and in a steady state over the time-scales of interest, even in the presence of radiative cooling and thermal conduction. Thus, the unperturbed state can be described by the time-independent axisymmetric pressure $P_0$, density $\rho_0$, temperature $T_0$ and velocity field $\bfv_0 = (v_{0R}, v_{0z}, v_{0\phi})$, which satisfy Eqs. \eqref{eq:continuity}-\eqref{eq:entropy_cons} with vanishing partial derivatives with respect to $t$. 

We linearise Eqs. \eqref{eq:continuity}-\eqref{eq:entropy_cons} with Eulerian perturbations of the form $F_0+\delta F\exp\left(ik_R R + ik_z z -i\omega t\right)$, where $F_0$ is the equilibrium value and $\left|\delta F\right|\ll \left|F_0\right|$. We work in the short-wavelength (WKB) limit, i.e. we assume that $kH \gg 1$ where $H = \cs^2/|\bfg_{\rm eff}|$ is the typical lengthscale of the system. We also use the Boussinesq approximation, which allows to filter out sound wave modes from the dispersion relation so that only the relevant low-frequency modes with $\omega \ll | \bfk | c_{\rm s}$ remain (see for example \citealt{Nipoti2010,BalbusPotter2016}).

Under these assumptions one finds the following dispersion relation:
\begin{equation}
\label{eq:DR}
\tilde{\omega}^3 + i \omega_{\rm d} \tilde{\omega}^2 - \left(\omega_{\rm BV}^2 + \omega_{\rm rot}^2\right) \tilde{\omega} - i \omega_{\rm rot}^2\omega_{\rm d} = 0 \;,
\end{equation}
where $\tilde{\omega} = \omega-k_R v_{0R}-k_z v_{0z}$, so that a mode is exponentially growing and thus formally unstable if and only if $\operatorname{Im}(\tilde{\omega})>0$. Equation \eqref{eq:DR} is formally identical to the dispersion relation obtained in the local analysis of Paper {\sc I} and therefore confirms that we can use the results obtained in that paper for the problem considered here. Indeed, the problem studied in Paper {\sc I} can be considered a simplified version of the problem studied by \cite{Nipoti2010} that makes the following additional assumptions (not present in Nipoti's paper): (i) no differential rotation; (ii) the surfaces of constant pressure and density coincide. 

In the above we have defined the following characteristic frequencies:

\begin{align}
& \omega_{\rm BV}^2  = -\frac{k_z^2}{k^2}\frac{\mathcal{D}\left( P_0 \right)}{\gamma\rho_0}\mathcal{D} \left( \sigma_0 \right) \,, \label{eq:omegaBV}  \\
& \omega_{\rm rot}^2  = -\frac{k_z^2}{k^2}\frac{1}{R^3} \mathcal{D}\left(R^4\Omega^2\right) \;, \label{eq:omegarot} \\
& \omega_{\rm d}  = \omega_{\rm c}+\omega_{\rm th} \;.
\end{align}
where
\begin{align}
& \mathcal{D}= \frac{k_R}{k_z}\frac{\pa}{\pa z}-\frac{\pa}{\pa R}\,, \\
& \omega_{\rm c} = \frac{\gamma-1}{\gamma} \frac{k^2 f\kappa_{\rm S} T_0^{7/2}}{P_0} \,, \\
& \omega_{\rm th} = -\frac{\gamma-1}{\gamma} \left(\frac{\rho_0}{\mu m_{\rm p}}\right)^2 \frac{\Lambda\left(T_0\right)}{P_0}\Delta\left(T_0\right) \,, \\
& \Delta\left(T\right)= 2-\frac{{\rm d}\log\Lambda\left(T\right)}{{\rm d}\log T}\,,
\end{align}
and $\Omega = v_{0\phi}/R$.

\begin{figure*}
\centering
\includegraphics[width=0.95\textwidth]{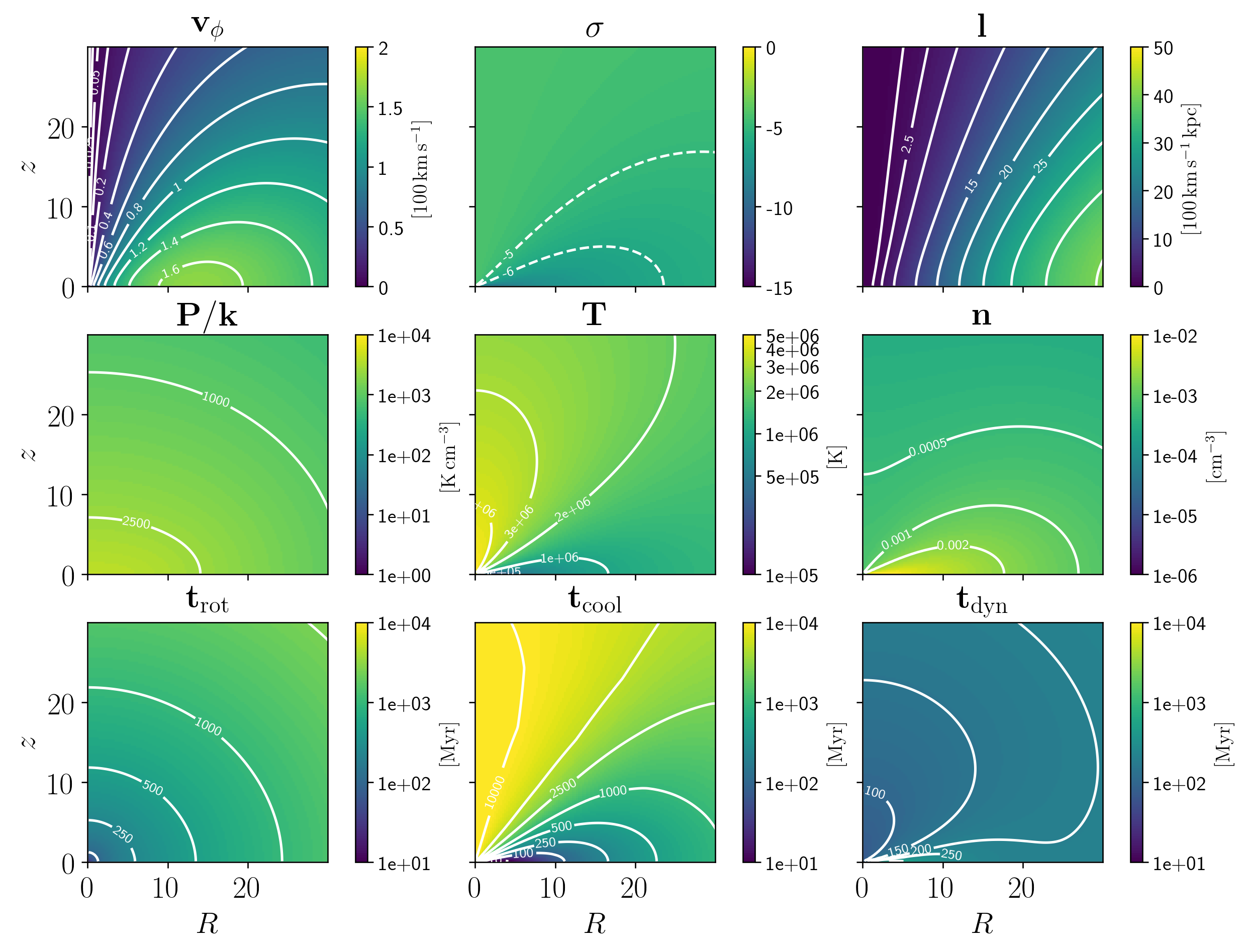}
\caption{The fiducial model of \citet{Sormani+2018c} (their model 6), which also serves as initial conditions for the simulations shown in Fig. \ref{fig:2} and \ref{fig:2bis}.}
\label{fig:1}
\end{figure*}

The quantity $\omega_{\rm BV}$ is the characteristic frequency of buoyant oscillations, while $\omega_{\rm rot}$ is the characteristic frequency related to the rotation of the system. Note that $\omega_{\rm BV}$ is related to the dynamical time \eqref{eq:tdyn} by $\omega_{\rm BV} \sim t_{\rm dyn}^{-1}$ and similarly $\omega_{\rm rot} \sim t_{\rm rot}^{-1}$ (the exact relations depend on the direction of ${\bf k}$). The combination $\omega_{\rm BV}^2 + \omega_{\rm rot}^2$ is positive when the equilibrium is convectively stable, as we assume in this paper,\footnote{The condition that $\omega_{\rm BV}^2 + \omega_{\rm rot}^2>0$ for all the possible $\left(k_R,k_z\right)$ gives the classical Solberg-H{\o}iland stability criteria \citep[e.g.][]{Tassoul2000}.} while the individual terms $\omega_{\rm BV}^2$ and $\omega_{\rm rot}^2$ are not necessarily positive. The quantity $\omega_{\rm d}$ is the characteristic frequency of dissipative processes, and is composed of the sum of the cooling ($\omega_{\rm th}$) and thermal conduction ($\omega_{\rm c}$) frequencies. We note $\omega_{\rm th}$ is negative if the plasma is thermally unstable in the classical sense of \cite{Field1965} (as we assume throughout this paper) while $\omega_{\rm c}$ is always positive, so the effect of thermal conduction always goes in the direction of stabilising the system (see below).

\begin{figure*}
\centering
\includegraphics[width=0.99\textwidth]{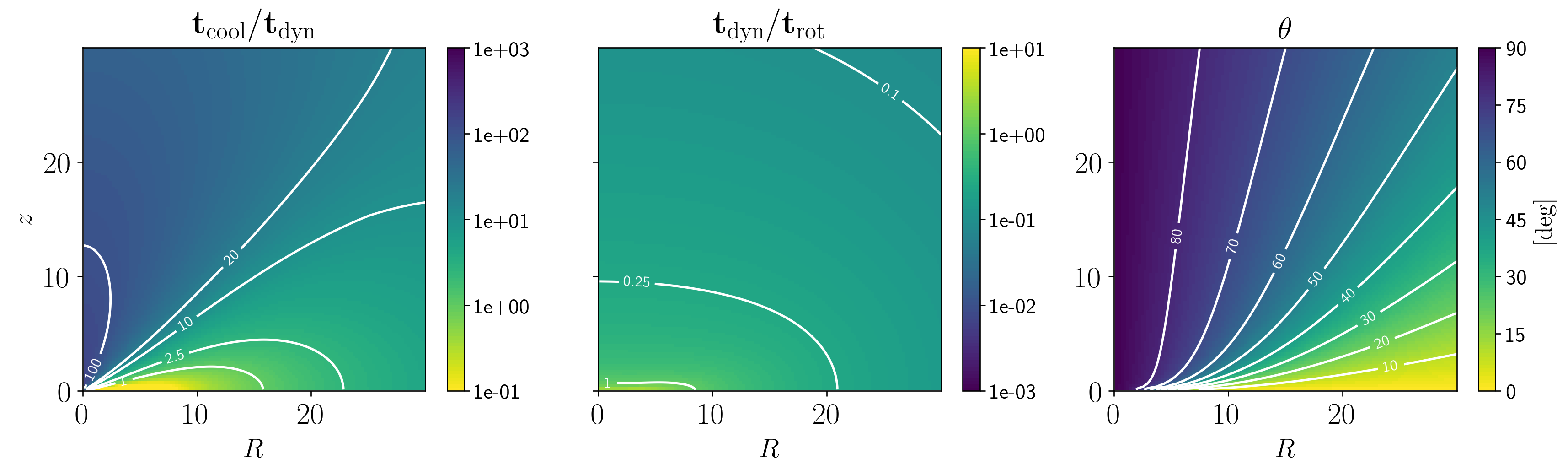}
\caption{The three important parameters that determine the thermal stability of the system (see Sect. \ref{sec:dr_analysis}) for the fiducial model shown in Fig. \ref{fig:1}. Condensations are expected where $t_{\rm cool}/t_{\rm dyn}\lesssim 10 f_{\rm rot} $, where $f_{\rm rot} = f_{\rm rot}(t_{\rm dyn}/t_{\rm rot}, \theta) \geq 1$ is the factor by which condensations are enhanced by the presence of rotation (see fig. 4 of Paper {\sc I}).}
\label{fig:1bis}
\end{figure*}

\subsubsection{Analysis of the dispersion relation}

Since the dispersion relation \eqref{eq:DR} is formally identical to the one derived in Paper {\sc I}, we will not repeat the analysis here but simply state some points that are relevant for this paper.

\begin{itemize}
\item In the fast cooling regime ($t_{\rm cool} \ll$ all other timescales so that we can set $\omega_{\rm BV}=\omega_{\rm rot} = \omega_{\rm c} = 0$), we recover the classical \citet{Field1965} instability, and the dispersion relation reduces to
\begin{equation}
\tilde{\omega} = - i \omega_{\rm th} \;.
\end{equation}
Hence, condensations are expected in this regime.
\item As $t_{\rm cool}$ (and in particular the ratio $t_{\rm cool}/t_{\rm dyn}$) is increased, the number of unstable modes decreases (i.e. the modes corresponding to more and more $\mathbf{k}$ turn into overstable modes). For example, in the limit $t_{\rm cool}/t_{\rm dyn} \gg 1$ in the non-rotating case only the modes with ${\bf k}$ nearly parallel to $\nabla P_0$ remain unstable ((in the non-rotating case this means that ${\bf k}$ is radial, so that $\omega_{\rm BV}=0$ for these modes, see also \citealt{Binney+2009}). This suggests that the instability becomes weaker and weaker as $t_{\rm cool}/t_{\rm dyn}$ is increased and leads to the expectation that there may exist a threshold value of $t_{\rm cool}/t_{\rm dyn}$ above which condensation cannot occur.
\item A new unstable mode appears due to the presence of rotation. This was already pointed out by \cite{Nipoti2010}. In the slow cooling regime ($t_{\rm cool}/t_{\rm dyn} \gg 1$), neglecting the effect of thermal conduction ($f=0$), this mode grows with a frequency of (see Paper {\sc I}):
\begin{equation} \label{eq:omega_new}
\tilde{\omega} = - i \frac{\omega_{\rm rot}^2}{\omega_{\rm BV}^2 + \omega_{\rm rot}^2} \omega_{\rm th}.
\end{equation}
Since $\omega_{\rm BV}/\omega_{\rm rot} \sim t_{\rm rot} / t_{\rm dyn}$, this mode becomes important when $t_{\rm dyn}/t_{\rm rot}$ is large (note that in this limit $\tilde{\omega} \sim - i \omega_{\rm th}$), i.e. when the corona has substantial rotational support. We discuss the physical interpretation of this new mode in Appendix \ref{sec:interpretation}.
\item When thermal conduction is added to the system, the dispersion relation is formally identical with the substitution $\omega_{\rm th} \to \omega_{\rm d}$. Thus the effect of thermal conduction is to stabilise all modes with wavelengths shorter than:
\begin{equation} \label{eq:lambdacrit}
\lambda_{\rm crit} = \frac{2 \pi \mu m_{\rm p} }{\rho_0} \left[ \frac{f \kappa_{\rm S} T_0^{7/2}}{\Lambda(T_0) \Delta(T_0)} \right]^{1/2}.
\end{equation}
This condition is obtained by requiring that $\omega_{\rm d}=0$.
\end{itemize}

\subsection{When do we expect condensations?} \label{sec:dr_analysis}

From the linear analysis it follows that (in the absence of thermal conduction) condensation will certainly occur in the fast cooling regime $t_{\rm cool}/t_{\rm dyn} \ll 1$ (because we recover the classical \citealt{Field1965} instability, see previous section). Considerations that use the linear analysis as a guide to infer what happens in the non-linear regime lead to the expectation that condensations should not occur in the opposite regime $t_{\rm cool}/t_{\rm dyn} \gg 1$ \citep[][see also Sect 3.1.2 and 4.3 of Paper {\sc I}]{Malagoli+1987, Binney+2009}.\footnote{Intuitively, the physical reason why condensations do not form if $t_{\rm cool}/t_{\rm dyn} \gg 1$ essentially boils down to the fact that an overdense clump will sink and mix with the hot gas before having time to cool significantly.} Hence there should exist a threshold value of $t_{\rm cool}/t_{\rm dyn}$ above which condensation cannot occur. While the linear analysis constitutes a useful guide to what we can expect in the non-linear regime, the physics involved is inherently non linear, so the exact value of $t_{\rm cool}/t_{\rm dyn}$ above which condensations cannot occur cannot be inferred from the linear analysis but must be determined using numerical simulations (\citealt{McCourt+2012}, Paper {\sc I}).

In the absence of rotation ($t_{\rm dyn}/t_{\rm rot}=0$), \citet{McCourt+2012} and \citet{Sharma+2012} suggested that condensation will occur when $t_{\rm cool}/t_{\rm dyn}\lesssim 10$, the threshold being quite sharp.\footnote{\citet{McCourt+2012} found a somewhat more stringent criterion, $t_{\rm cool}/t_{\rm dyn}\lesssim 1$, than \cite{Sharma+2012}, who found $t_{\rm cool}/t_{\rm dyn}\lesssim 10$. Previous claims attributed the difference to the different geometry (plane parallel vs spherical), but \citet{ChoudhurySharma2016} has shown that the difference is due to the fact that \citet{McCourt+2012} looked at the condensation at the location of $\min(t_{\rm cool}/t_{\rm ff})$, while \citet{Sharma+2012} considered condensations anywhere in the box, which happens further inside the location of the initial $\min(t_{\rm cool}/t_{\rm ff})$. Also note that the threshold becomes even less stringent if large density fluctuations are assumed to be present in the initial conditions \citep[e.g.][]{Choudhury+2019}.}

In Paper {\sc I} we have found that the presence of rotation can increase the threshold value of $t_{\rm cool}/t_{\rm dyn}$ by a factor up to $f_{\rm rot} \sim 10$. The effect becomes largest when (i) rotation is dynamically important (namely, in the regime $\Omega c_{\rm s}/g_{\rm eff}\gtrsim 1$, or equivalently $t_{\rm dyn}/t_{\rm rot}\gtrsim 0.2$); (ii) $\theta$ is small, where $\theta$ is defined as the angle between ${\bf g}_{\rm eff}$ and the direction perpendicular to ${\pmb\Omega}$ (see fig. 1 of Paper {\sc I}), which occurs preferentially near the Galactic plane. The strength of this effect as a function of the parameters is quantified in fig. 4 of Paper {\sc I}. 

While item (i) could be interpreted as a result of the new unstable mode that appears as a result of rotation (Eq. \ref{eq:omega_new}), the fact that condensation depends on $\theta$ (item ii) must be an entirely non linear effect because the dispersion relation \eqref{eq:DR} does not depend on $\theta$. In Paper {\sc I}, we have interpreted this as a result of the fact that the motion in the direction perpendicular to ${\pmb\Omega}$ is hampered by the conservation of the angular momentum, so contrary to the non-rotating case an overdense region will not sink and mix if ${\bf g}_{\rm eff}$ and ${\pmb\Omega}$ are nearly perpendicular (see Sect. 4.3 in Paper {\sc I}).

In conclusion, we may expect the thermal instability in a rotating corona in the absence of thermal conduction to occur when $t_{\rm cool}/t_{\rm dyn} \lesssim 10 f_{\rm rot}$ where $f_{\rm rot} \geq 1$ is the factor by which the threshold is increased by the presence of rotation, which depends on $t_{\rm dyn}/t_{\rm rot}$ and $\theta$ (see fig. 1 of Paper {\sc I} for the definition of $\theta$). The factor $f_{\rm rot}$ can be read off the right panel in fig. 4 of Paper {\sc I}. The presence of thermal conduction has a stabilising effect up to a certain maximum spatial scale given by Eq. \eqref{eq:lambdacrit}.

\subsection{Application to the equilibrium models of \citet{Sormani+2018c}} \label{sec:linearapplication}

We now apply the considerations of the previous sections to the models of \citet{Sormani+2018c}, to see where whether we may expect them to produce condensations. 

Fig. \ref{fig:1} shows the fiducial model of \citet{Sormani+2018c}. According to Sect. \ref{sec:dr_analysis}, the parameters that control the thermal stability of the system are $t_{\rm cool}/t_{\rm dyn}$, $t_{\rm dyn}/t_{\rm rot}$ and $\theta$ (the angle between the effective gravity and the direction perpendicular to the rotation axis, see fig. 1 of Paper {\sc I}). These are shown in Fig. \ref{fig:1bis}. The regions where the condensations are most likely to happen are those where $t_{\rm cool}/t_{\rm dyn}$ becomes small, $t_{\rm dyn}/t_{\rm rot}$ becomes large and $\theta$ becomes small. Hence these are the regions we should look for. Excluding the region $z \lesssim 5\kpc$, which is the transition zone where the corona is interacting with the Galactic disc (of which our model cannot be taken as a faithful representation), we see that the most favourable area is the region $5 \lesssim z \lesssim 10\kpc$, $10 \lesssim R \lesssim 30 \kpc$. Interestingly, this roughly coincides with the locations where HVCs are found in the MW: as mentioned in the introduction, most of the HVCs are found in proximity of the Galactic disc, at heights $z \lesssim 10 \kpc$, distances from the Sun between $2 < d <15 \kpc$ and within $30\degree$ from the disc plane as viewed from the Galactic centre. Note that, in contrast, in a non-rotating spherical model the most favourable region for the formation of the HVCs (for example, model 1 or 4 of \citealt{Sormani+2018c} or the models discussed in \citealt{Binney+2009}) is a spherical region within few kpc of the Galactic centre, in disagreement with observations.

We see from Fig. \ref{fig:1} that the presence of rotation is expected to help the production of condensation in the favourable area mentioned above by lowering the threshold $t_{\rm cool}/t_{\rm dyn}$ by a factor of $f_{\rm rot}\sim$ a few according to the right panel of fig. 4 of Paper {\sc I}. This is a \emph{direct} effect by which condensations are enhanced by rotation.

\begin{figure}
\centering
\includegraphics[width=0.45\textwidth]{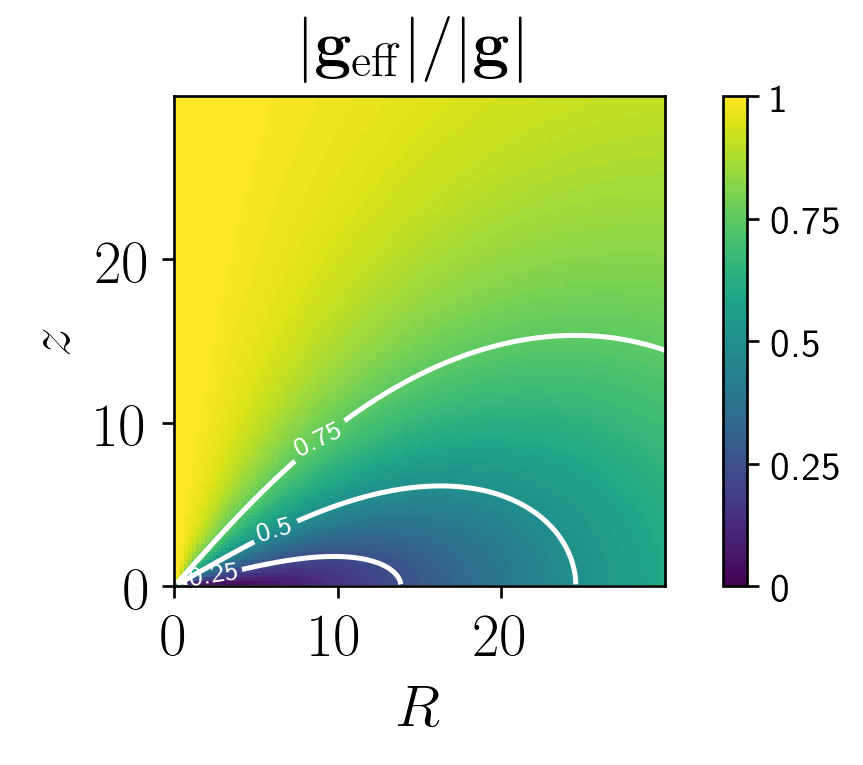}
\caption{The ratio between $\bfg_{\rm eff} = - \nabla \Phi + \Omega^2 R \; \hateR$ and $\bfg = - \nabla \Phi$ for the fiducial model shown in Fig. \ref{fig:1}.}
\label{fig:3}
\end{figure}

There is also an \emph{indirect} effect that makes condensation more likely in a rotating corona compared to a non-rotating corona in the same gravitational potential. Fig. \ref{fig:3} shows the ratio $|\bfg_{\rm eff}| /| \bfg |$ for the same model as in Fig. \ref{fig:1}. From this figure we see that compared to a non-rotating corona in the same external gravitational potential (i.e. with the same value of $\mathbf{g}$) and with the same density and temperature, the presence of rotation increases the value of $t_{\rm dyn}$ (Eq. \ref{eq:tdyn}) by roughly a factor of 2 in the relevant regions, and therefore reduces the ratio $t_{\rm cool}/t_{\rm dyn}$ (which is the parameter that matters for the production of condensations) by the same factor. This makes condensations more likely.

The fact that the formation of HVCs is enhanced in a region in proximity of the Galactic plane ($z \lesssim 10\kpc$) at radial distances of few tens of $\kpc$ is not a particular feature of our fiducial model, but is a quite a general feature of any plausible rotating model. Any model which rotates at a substantial fraction of the circular velocity of the disc will generate (i) a similar enhancement in density and therefore of $t_{\rm dyn}/t_{\rm cool}$ in the relevant regions, and (ii) a relative enhancement of $T$ in a cone above the Galactic centre (i.e. whose axis is $R=0$) compared to a spherical model, which strongly suppresses the formation of HVCs in this region. Indeed, other models such as for example the models of \cite{Pezzulli+2017} and all the rotating models in \citet{Sormani+2018c} have similar characteristics.

\begin{figure}
\centering
\includegraphics[width=0.45\textwidth]{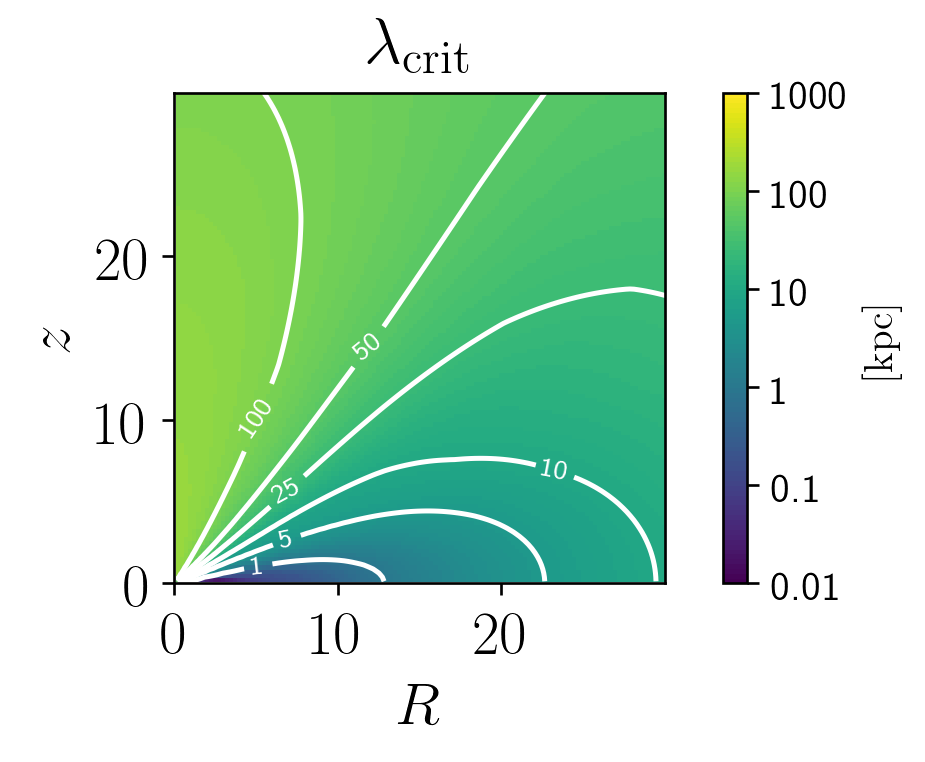}
\caption{Critical wavelength below which modes are stabilised by thermal conduction (see Eq. \ref{eq:lambdacrit}) for the fiducial model shown in Fig. \ref{fig:1}.}
\label{fig:4}
\end{figure}

Fig. \ref{fig:4} shows the critical wavelength below which modes are stabilised by thermal conduction for the same model, assuming that the suppression factor of thermal conduction due to a tangled magnetic field is $f=0.01$ (the same value used in \citealt{Binney+2009}). This shows that thermal conduction is likely to be very effective in further suppressing the formation of condensations in a cone above the Galactic centre. In the relevant region where HVCs may be expected to form ($5 \lesssim z \lesssim 10\kpc$, $10 \lesssim R \lesssim 30 \kpc$), the situation is less clear. The figure suggests that suppression due to thermal conduction may be substantial but, given that the values of $\lambda_{\rm crit}$ are too large for the WKB approximation which underlies the linear analysis to be valid, it is unclear whether this will actually stop condensations from forming.

In conclusion, the analysis in this section suggests that condensation might form in a rotating corona due to a combination of the effects discussed in Sect. \ref{sec:dr_analysis}. It is unclear whether thermal conduction is able to suppress the formation of condensations. To determine this it is necessary to run some numerical experiments.

\section{Numerical experiments} \label{sec:nonlinear}

To better understand the non-linear behaviour of the thermal instability, we run axisymmetric 2.5D hydrodynamical simulations in the presence of cooling and thermal conduction using the models of \citet{Sormani+2018c} as initial conditions. We do not impose any initial perturbation. In these simulations, the evolution of the system is governed by Eqs. \eqref{eq:continuity}-\eqref{eq:entropy_cons}. Note that our initial conditions are not equilibria with respect to the energy equation (Eq. \ref{eq:entropy_cons}). These simulations are 2.5D in the sense that we assume axisymmetry, so the grid in the $(R,z)$ plane is two-dimensional, but the velocity vector has three components $(v_R,v_z,v_\phi)$.

\subsection{Numerical setup}

We have used the public Eulerian code {\sc PLUTO} version 4.3 \citep{Pluto2007}. This is a free software for the numerical solution of systems of conservation laws targeting high Mach number flows in astrophysical fluid dynamics. We use a two-dimensional static cartesian grid in the region $R \times z = [0,200] \kpc \times [0, 200] \kpc$. The grid is uniformly spaced in both $R$ and $z$ with $800 \times 800$ points, which corresponds to a spatial resolution of $0.25 \kpc$. We use the following parameters: {\sc rk2} time-stepping, no dimensional splitting, {\sc roe} Riemann solver and the {\sc mc} flux limiter. Boundary conditions are reflective on the plane $z=0$ and on the outer boundaries at $z=200\kpc$ and $R=200\kpc$ and axisymmetric (i.e. $v_{\phi}$ is reversed on the ghost cells just beyond the axis $R=0$) on the axis $R=0$. Thermal conduction is treated with the {\sc Explicit} scheme. 

We adopt the collisional ionisation equilibrium cooling function from \cite{sd93} for ${\rm [Fe/H]}=-0.5$. This accounts for the sub-solar metallicities found in the HVCs, which are believed to characterise the Galactic corona \citep[e.g.][]{VanWoerdenWakker2004,Bregman+2018}. The cooling is switched off for $T\leq 10^4 \, \rm K$, and for $z \leq z_0=5\kpc$ since in this region the corona is likely interacting with the Galactic disc and our model cannot be taken as a faithful representation of the real physical conditions expected in this region. Indeed the cooling timescale of our model close to the plane becomes very short and the presence of cooling would lead to a cooling catastrophe. To make the transition smoother, for $z>z_0$ we add a position-dependent and time-independent heating function $H(R,z) = \exp[-(z - z_0)/z_0] \, H_0(R,z)$ where the function $H_0(R,z)$ is such that at $t=0$ the heating exactly balances the cooling. We have experimented with different choices for the way in which the cooling is switched off near to the plane, and all lead to similar results (see discussion in Sect. \ref{sec:discussion}).

Finally, we have verified by running simulations in which both cooling and thermal conduction are switched off that all the models of \citet{Sormani+2018c} are indeed stable equilibrium models. In this setup, the initial state does not change noticeably during the course of the simulations, which lasts for several Gyr.

\subsection{Results} \label{sec:sim_results}

\begin{figure*}
\centering
\includegraphics[width=1.0\textwidth]{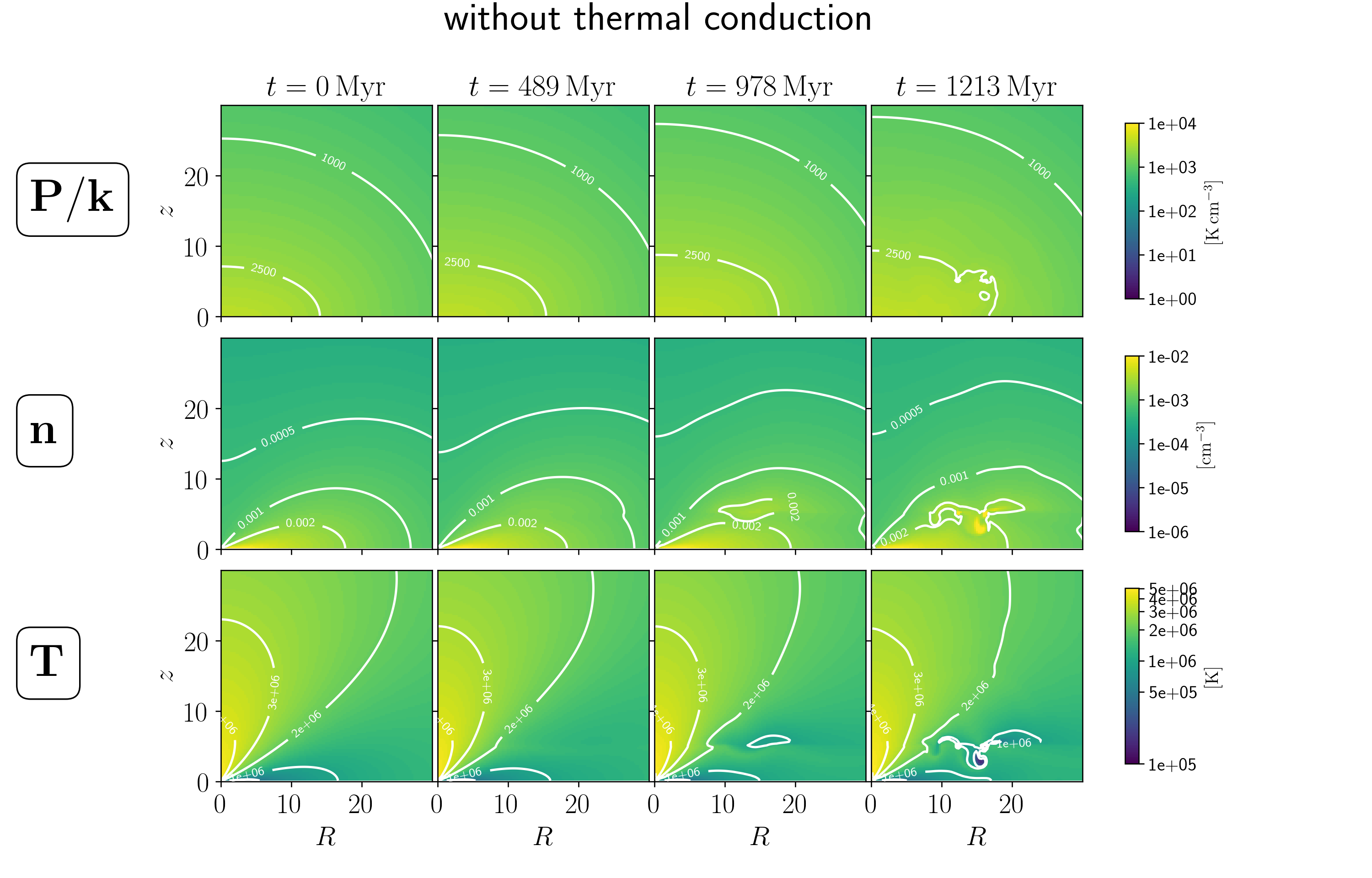}
\caption{Evolution of the fiducial model of \citet{Sormani+2018c} under the presence of the cooling function from \citet{sd93} for $\rm [Fe/H]=-0.5$ but in the absence of thermal conduction. }
\label{fig:2}
\end{figure*}

Fig. \ref{fig:2} shows the evolution of our fiducial model under the presence of cooling but in the absence of thermal conduction. The model slowly contracts as the consequence of cooling until at $t\simeq1 \Gyr$ a condensation forms at $z\sim6\kpc$, $R\sim16\kpc$. This leads to a big drop of condensed gas which falls onto the Galactic disc. This is quickly followed a sequence of similar drops which fall on the Galactic disc in a way reminiscent of a leaky faucet. The location where the condensation forms is in agreement with the expectations from the linear analysis (Sect. \ref{sec:linearapplication}) and is roughly consistent with the observationally determined locations of HVCs. This suggests that indeed thermal instability can produce condensation resembling HVCs. We discuss in more detail in Section \ref{sec:discussion} the dependence of these results on our assumptions, and in particular on our parametrisation of the heating and cooling functions at the disc-halo interface.

\begin{figure*}
\centering
\includegraphics[width=1.0\textwidth]{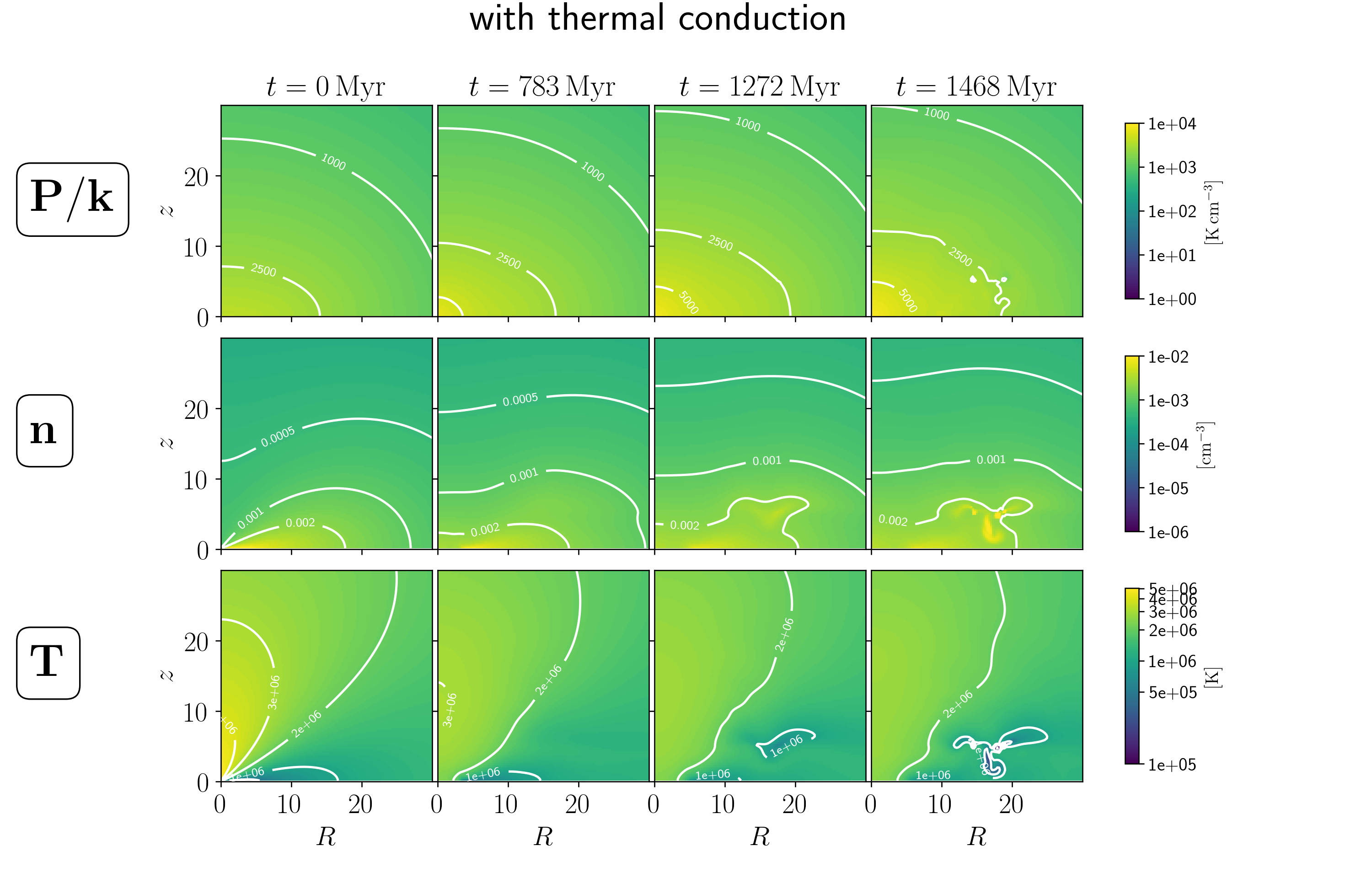}
\caption{Evolution of the fiducial model of \citet{Sormani+2018c} under the presence of the cooling function from \citet{sd93} for $\rm [Fe/H]=-0.5$ and thermal conduction.}
\label{fig:2bis}
\end{figure*}

Fig. \ref{fig:2bis} shows the evolution of the same model under the presence of both cooling and thermal conduction. The latter has the effect of slightly delaying the formation of the condensation and to slightly smooth out its density and temperature contrast \citep[for a detailed study of the dynamics of cold clouds in the presence of thermal conduction, see e.g.][]{Armillotta+2017}. However, it seems that thermal conduction is not able to halt the formation of the condensation. It must be mentioned however that this relies on a crude model for the suppression of thermal conduction due to the presence of a tangled magnetic fields and on the value of $f=0.01$. A proper treatment of the problem would require magneto-hydrodynamics rather than hydrodynamic simulations. We have tested that in a simulation with $f=1$ thermal conduction is so strong that not only suppresses the condensation, but also changes the entire structure of the corona making it almost isothermal within a few hundred of Myr.

We have tested what happens to all the other models presented in \citet{Sormani+2018c}. All the rotating models are unstable and produce condensations. The instability is much stronger in the isothermal rotating models (models 2 and 3) on account of their higher densities in proximity of the Galactic plane, while it is similar to our fiducial model in the adiabatic rotating model 5. The non-rotating spherical adiabatic model 4 is stable, and produces no condensations, as expected since its cooling times are very long everywhere \citep[$t_{\rm cool}/t_{\rm dyn}\gtrsim 100$ - see also the similar model in the central column of fig. 5 of][]{Binney+2009}. In the non-rotating spherical isothermal model 1 cooling times are short within a few kpc of the Galactic centre compared to the duration of the simulations \citep[similarly to the model in the central column of fig. 4 of][]{Binney+2009}, thus the models cools very quickly near the centre, producing condensations. However, these condensation are all within a roughly spherical regions of $r \lesssim 8 \kpc$ from the Galactic centre, so they are simply a central accumulation of cold gas and do not agree with the observationally determined locations of HVCs.

\section{Discussion}
\label{sec:discussion}

The results presented in Sections \ref{sec:linearapplication} and \ref{sec:sim_results} suggest that condensation may form in a rotating stratified corona. However, there are some aspects that deserve further discussion.

The first aspect is the turn off of the cooling near the Galactic plane for $z\leq z_0$. We have experimented with different values of $z_0$ and with different ways of turning off the cooling close to the Galactic plane. The latter always produce similar results for comparable values of $z_0$. For example, we can (i) turn off the cooling gradually following a function $F(z) = \tanh[(z-z_0)/z_1]$, where $z_0$ and $z_1$ are two parameters, instead of using a heating function as in Sect. \ref{sec:nonlinear}; (ii) assume that heating balances the cooling on large scales (the so called ``global thermal balance'') if $z<z_0$, and turn off the cooling at $z<z_1$. This produces results that are very similar to those described in Sect. \ref{sec:sim_results} for comparable values of $z_0$ and, say, $z_1 = 2 \kpc$ (results are not very dependent on this value, provided that $z_1\ll z_0$). The effects of changing the value of $z_0$ are more tricky. If $z_0$ is increased to, say, $z_0=7.5\kpc$, then condensations will form at higher $z$ and at later times than shown in Fig. \ref{fig:2}. If we raise $z_0$ to even higher values, say $z_0=20\kpc$, condensations will stop altogether forming because the cooling time becomes comparable to the Hubble time at those heights. The trickiest point is that, for example, while condensations form at $z\gtrsim 7.5\kpc$ if $z_0=7.5\kpc$, they will form at lower heights if $z_0$ is lower. This is because they form first where the cooling time is shortest, and the upper parts of the atmosphere then collapse before having time to produce condensations. If $z_0\sim 0$ our setup reduces to a pure cooling flow and no condensations are observed outside of the disc: the gas is falling too fast and collapses onto the disc before having time to cool significantly. We may argue that near the Galactic plane some form of heating should be present, and that excessive cooling in the corona at $z \lesssim 5\kpc$ must be somehow prevented, and thus condensations should form at those heights. We conclude that condensations resembling HVCs may form only if the corona is well represented by a nearly-equilibrium stratified rotating structure near the plane, while they cannot form if it is well-represented by a fast cooling flow. The detailed study of this question requires better understanding of the conditions in the transition zone between the disc and the Galactic atmosphere, which is a complicate and poorly understood problem at the moment.

The second aspect is that it is difficult to disentangle which of the two effects discussed in Sect. \ref{sec:linearapplication} is actually responsible for the production of condensations. The presence of rotation enhances the condensations (i) indirectly, by affecting the global density and temperature structure of the corona and thus changing the global profile of $t_{\rm cool}/t_{\rm dyn}$, which is the parameter that controls the formation of condensations; (ii) directly, by increasing the threshold value of $t_{\rm cool}/t_{\rm dyn}$ below which condensations can form. However, the locations where these two effects become relevant are almost identical in the models. We conclude that it is probably a combination of both effects that is at work here and makes the corona thermally unstable. Note, however, that effect (ii) relies on the results of Paper {\sc I} which does not include the effect of differential rotation, whose impact on condensations is yet to be explored in the non-linear regime.

It should also be mentioned that throughout this paper we assume axisymmetry, but we cannot exclude a priori that the formation of condensation is suppressed when this assumption is relaxed \citep[][]{Nipoti2010,NipotiPosti2014}. To check this would require 3D simulations, which are however much more computationally expensive than the ones presented in this paper.

The pressure within the condensation in Fig. \ref{fig:2} and \ref{fig:2bis} is slightly lower than in the surrounding hot medium. This is in agreement with the fact that the measured pressures of HVCs are slightly below what is expected from pressure equilibrium with the hot medium, where the pressure of the hot medium is based on its density and temperature determinations (see for example fig. 1 and remarks in Sect. 4.2 in \citealt{Sormani+2018c}).

We have crudely modelled the thermal conduction by taking into account its suppression due to a tangled magnetic field using a factor $f$. However, we note that this problem should be properly studied in the context of magnetohydrodynamics. This lies outside of the scope of the current paper, but is a worthwhile direction for future investigations. Indeed, magnetic field are known to have non trivial effects on the dynamics of cold gas clouds travelling through the hot coronal medium \citep[e.g.][]{Gronnow+2017, Gronnow+2018}.

Finally, since density determinations of the Galactic corona have large uncertainties, we have experimented by running simulations of our fiducial model in which $T_0(R,z)$ and $v_{0 \phi}(R,z)$ are unchanged but the density is scaled down by a factor of 2. This lowers the cooling time and thus has the effect of delaying the formation of condensation by approximately the same factor. Hence condensations may be suppressed only if the density is lower by a factor of $\sim 10$, which would lead to condensation forming on times longer than the Hubble time. However, based on current constraints on the density of the Galactic corona this seems unlikely albeit possible. In relation to this it is worth mentioning that, while in this paper we have assumed that the densities and pressure of the corona are a `given background', they may be set by the requirement that the gas the value of $t_{\rm cool}/t_{\rm dyn}$ is close to the threshold value for the production of condensations \citep[e.g.][]{Sharma+2012b,Voit+2015,Voit+2019}. The consequences of this requirement are yet to be explored in the context of rotating coronae.

\section{Conclusions}
\label{sec:conclusions}

We have examined the effects of rotation on the formation of HVCs in the Galactic corona. We have used both a linear stability analysis, and the results of local and global hydrodynamical simulations applied to the models of \citet{Sormani+2018c}.

Rotation enhances the formation of condensation through two effects:
\begin{enumerate}
\item An \emph{indirect} effect. Rotation changes the global density and temperature structure of the corona, thus changing the global profile of $t_{\rm cool}/t_{\rm dyn}$, the parameter that controls the formation of condensations (see Sect. \ref{sec:linearapplication} and Figs. \ref{fig:1}, \ref{fig:1bis} and \ref{fig:3}).
\item A \emph{direct} effect. The presence of dynamically significant rotation ($\Omega c_{\rm s}/g_{\rm eff}\gtrsim 1$ or equivalently $t_{\rm dyn}/t_{\rm rot} \gtrsim 0.2$) enhances the condensation by directly changing the threshold value of $t_{\rm cool}/t_{\rm dyn}$ below which condensations can form. The effect is largest in the proximity of the Galactic plane, where the effective gravity is nearly perpendicular to the rotation axis (see Sect. \ref{sec:dr_analysis} and \ref{sec:linearapplication} and fig. 4 in Paper {\sc I}). This is the effect we discovered in Paper {\sc I} and assumes that we can neglect the impact of differential rotation, whose importance for the condensation remains to be tested in the non-linear regime.
\end{enumerate}

It is difficult to disentangle the two effects because the locations where they are important are essentially the same. In any reasonable \emph{rotating} model of the Galactic corona this makes the formation of condensation more favourable precisely in the location where HVCs are found in the observations (see Sect. \ref{sec:linearapplication}). Although this may be a coincidence and it may be that current observations are missing some neutral gas further out, given that several independent lines of research point to a corona that is significantly rotating in the inner parts (see Introduction) at the moment it seems reasonable to interpret this accordance as evidence that HVCs are formed by the thermal instability.

Thermal conduction does not suppress the formation of condensations, assuming that it is lower by a factor $\lesssim 10^{-2}$ with respect to the Spitzer's standard value due to the presence of a tangled magnetic field. However this treatment is rather crude, and a proper study requires explicitly taking explicitly into account the often subtle effects of the magnetic fields which is outside the scope of this work.

Condensations at $z\gtrsim5 \kpc$ are absent in our simulations if we completely turn off any form of heating near the Galactic plane. In this case a fast cooling flow is established near the disc and the gas does not have time to cool significantly before falling onto the disc. Hence, the formation of HVCs requires some amount of heating (e.g. supernova feedback) that prevents excessive cooling near the disc. In this sense, our results indicate that the disc-corona interaction is important, similarly to what has been suggested by \cite{Fraternali+2015}. The main difference between \cite{Fraternali+2015} and the scenario presented here is that we do not require the condensations to be `seeded' by a powerful superbubble that reaches the corona, but can happen spontaneously. We suggest that the formation of HVCs occurs under much more general circumstances, and only require that the Galactic corona is approximately in global thermal balance relatively close to the Galactic disc.

We conclude that if the corona rotates:
\begin{enumerate}[(a)]
\item HVCs can occur as a consequence of the thermal instability in the Galactic corona. Our conclusion differs from that of \cite{Nipoti2010} because his analysis was limited to the linear regime, while the direct effect (ii) mentioned above is largely a consequence of non-linear effects (Paper {\sc I}), and because he did not apply his analysis to any concrete global model of rotating corona so he also partly missed the importance of the indirect effect (i).
\item The predicted locations for the formation of the condensations agrees with the observationally determined locations of most HVCs, which are observed at $z$-heights $\lesssim 10\kpc$, at distances between $2\mhyphen15\kpc$ from the Sun and within $30\degree$ from the disc plane as viewed from the Galactic Center \citep[e.g.][]{Putman+2012}. This does not require much fine tuning, and is a general property of models of the MW corona with a substantial rotational support.
\item The formation of HVCs requires the corona to be in global thermal balance  (namely, heating roughly balances cooling on average) close to the disc. Hence, further understanding of the properties of HVCs requires a detailed modelling of the disc-halo interface, including the effect of supernova feedback.
\end{enumerate}

\section*{Acknowledgements}

The authors thank Lucia Armillotta, Filippo Fraternali, Uri Keshet, Ralf Klessen, Federico Marinacci, Carlo Nipoti, Gabriele Pezzulli and Steve Shore for useful comments and discussions. We are grateful to the referee, Prateek Sharma, for constructive suggestions that improved the paper. MCS acknowledges support from the Deutsche Forschungsgemeinschaft via the Collaborative Research Centre (SFB 881) ``The Milky Way System'' (sub-projects B1, B2, and B8). ES acknowledges support from the Israeli Science Foundation (grant 719/14) and from the German Israeli Foundation for Scientific Research and Development (grant I-1362-303.7/2016).

\def\aap{A\&A}\def\aj{AJ}\def\apj{ApJ}\def\mnras{MNRAS}\def\araa{ARA\&A}\def\aapr{Astronomy \&
 Astrophysics Review}\def\apjs{ApJS}\def\apjl{ApJ}\def\pasj{PASJ}\def\nat{Nature}\def\prd{Phys. Rev. D}
\def\ssr{Space Sci. Rev.}\def\pasp{PASP}\def\pasa{Publications of the Astronomical Society of Australia}
\bibliographystyle{mn2e}
\bibliography{bibliography}

\appendix

\section{Physical interpretation of the unstable mode that arises due to rotation} \label{sec:interpretation}

In the absence of rotation, the modes described by the dispersion relation \eqref{eq:DR} essentially describe blobs undergoing buoyant oscillations while cooling is acting. When rotation is added to the system, as noted by \cite{Nipoti2010}, a new unstable mode appears, which in the slow cooling regime has a frequency given by Eq. \eqref{eq:omega_new}. What is the physical interpretation of this new mode? 

In the limit $t_{\rm cool} \to \infty$ the frequency of the new unstable mode $\omega \to 0$ (Eq. \ref{eq:omega_new}). Hence the perturbed state is just a new equilibrium state in this limit and fluid elements associated with this type of perturbations maintain their equilibrium values in the $(R,z)$ plane at all times. Hence, these modes are clearly not buoyant oscillations. Indeed, it can be shown from the linearised equations of motion that the perturbations associated with this new unstable mode, in the slow cooling regime ($t_{\rm cool} \to \infty$), have $\delta v_R \to 0$ and $\delta v_z \to 0$. We now show how such modes can be constructed explicitly in this limit.

\begin{figure}
\centering
\includegraphics[width=0.4\textwidth]{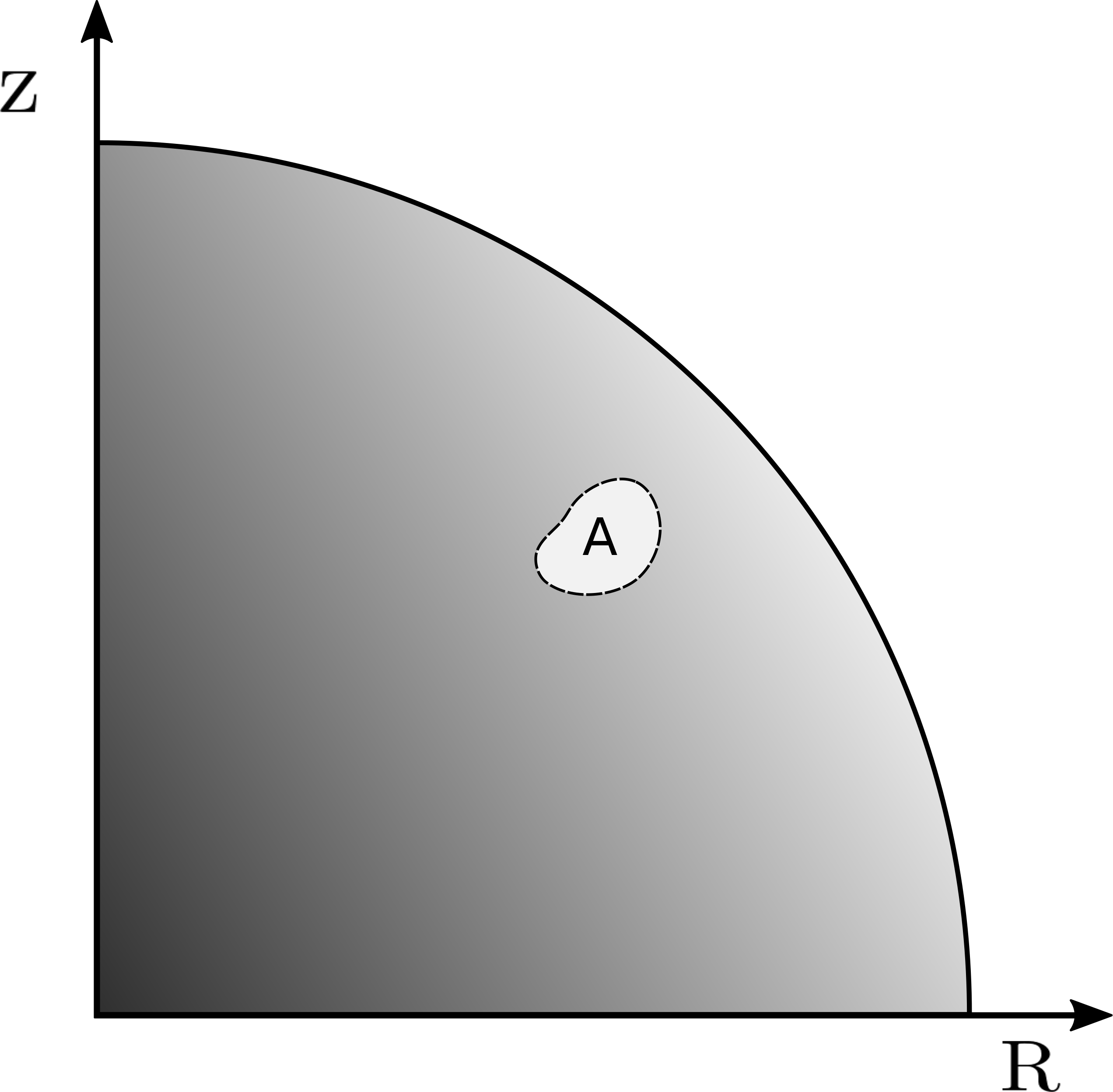}
\caption{Physical interpretation of the new unstable mode that appears due to rotation (see Appendix \ref{sec:interpretation}).}
\label{fig:interpretation}
\end{figure}

Imagine taking an equilibrium model such as those described in \citet{Sormani+2018c}. Can we find another equilibrium model which is exactly the same as the previous one everywhere except within a region $A$ (see Fig. \ref{fig:interpretation})?

Since as discussed in \citet{Sormani+2018c} an equilibrium is completely characterised by the pressure distribution $P_0(R,z)$, this is in principle easy to do: just take a function $P(R,z) = P_0(R,z) + \delta P(R,z)$ such that $P(R,z) = P_0(R,z)$ outside $A$ and $P(R,z)\neq P_0(R,z)$ within $A$. This will result in an equilibrium corona that is identical to the original one except within the region $A$. It is clear that if the associated density perturbation $\delta \rho / \rho_0$ is non-negligible then in the presence of cooling this region will cool at a substantially different rate than its surroundings (because cooling depends on the density as $\rho^2$), leading to a thermal instability. However, the perturbation analysis which underlies the derivation of the dispersion relation \eqref{eq:DR} requires that (i) $\delta F / F_0 \ll 1$ where $F$ is any of $P,\rho,v_{\rm \phi}$ (ii) perturbations to be quasi-isobaric so that $\delta P / P_0 \ll \delta \rho / \rho$ as a consequence of the Boussinesq approximation which filters out sound waves (iii)  $\delta P$ varies on short spatial scales, so that $\left|\nabla\log\delta P\right|\gg \left|\nabla\log P_0\right|$, as a consequence of the WKB approximation. Thus the question is, given a pressure perturbation $\delta P$ that satisfies these requirements, can we achieve a non-negligible density perturbation within the region $A$? Only if this can happen will the system be unstable according to the linear perturbation analysis that leads to Eq. \eqref{eq:DR}.

Given an equilibrium characterised by a pressure distribution $P\left(R,z\right)$ and an external potential $\Phi\left(R,z\right)$, the density $\rho(R,z)$ and azimuthal velocity $v_\phi(R,z)$ can be readily calculated using the following formulas \citep{Sormani+2018c}:
\begin{align}
\label{eq:rho}
&\rho = - \frac{\pa P/\pa z}{\pa\Phi/\pa z}\;, \\
\label{eq:vphi}
& v_\phi^2 = R\left(\frac{\pa \Phi}{\pa R} - \frac{\pa P/\pa R}{\pa P/\pa z}\frac{\pa\Phi}{\pa z}\right)  \;.
\end{align}

Writing $P(R,z) = P_0\left(R,z\right)+\delta P\left(R,z\right)$, substituting into \eqref{eq:rho} and using assumptions (i)-(ii) above we find:
\begin{equation}
\frac{\delta\rho}{\rho_0}\sim\frac{\pa \delta P/\pa z}{\pa P_0/\pa z}\sim\frac{\delta P}{P_0}\frac{\pa \log\delta P/\pa z}{\pa \log P_0/\pa z} \gg \frac{\delta P}{P_0}\;,
\end{equation}
which means that we may indeed regard the rapidly varying component of the solution as being isobaric, as required by the Boussinesq approximation. Substituting $P(R,z) = P_0\left(R,z\right)+\delta P\left(R,z\right)$ into Eq. \eqref{eq:vphi} and linearising we find
\begin{equation}
\frac{2v_{0\phi}\delta v_\phi}{R} = \frac{1}{\rho_0}\frac{\pa\delta P}{\pa R} - \frac{\delta\rho}{\rho_0^2}\frac{\pa P_0}{\pa R}\;.
\end{equation}
The ratio of the first to the second term on the right hand side is of order
\begin{equation}
\frac{1}{\delta\rho/\rho_0}\frac{\pa\delta P/\pa R}{\pa P_0/\pa R}\sim  \frac{\pa P_0/\pa z}{\pa \delta P/\pa z} \frac{\pa\delta P/\pa R}{\pa P_0/\pa R}\sim 1\;.
\end{equation}
Therefore, using the fact that $P_0\sim\rho_0 c_{\rm s}^2$ we see that
\begin{equation}
\frac{\delta v_\phi}{v_{0\phi}}\sim \frac{c_{\rm s}^2}{v_{0\phi}^2} \frac{\delta\rho}{\rho_0}\;.
\end{equation}
Putting everything together, we have that 
\begin{equation} \delta P / P_0 \ll \delta \rho / \rho_0 \sim (v_{0 \phi}^2 / c_{\rm s}^2) \delta v_{\phi} / v_{0\phi} \;.\end{equation}

Therefore, we can achieve a significant perturbation of $\delta \rho/\rho_0$ only if $(v_{0 \phi}^2 / c_{\rm s}^2) \gtrsim 1$, i.e. only if the corona rotates significantly. This explains why the unstable mode only becomes important when the corona is rotating significantly. If the corona does not rotate significantly instead, $(v_{0 \phi}^2 / c_{\rm s}^2) \ll 1$, the density perturbation $\delta \rho/\rho_0$ is suppressed by a factor $(v_{0 \phi}^2 / c_{\rm s}^2)$ compared to the velocity perturbation. Therefore, if the velocity perturbation is small as required by the linear regime, the density perturbation is negligible, which also explains why the frequency of the unstable mode given by Eq. \eqref{eq:omega_new} goes to zero as the rotation velocity $\Omega \to 0$. 

In conclusion, the new unstable mode that appears due to the fact that only in the presence of substantial rotational support it is possible to find a new equilibrium through a nearly-isobaric perturbation which has a non-negligible density perturbation. In the limit in which the corona is not rotating instead, a nearly-isobaric perturbation cannot achieve a substantial density perturbation unless the velocity perturbation $\delta v_\phi / v_{0\phi}$ is large, which is unlikely to happen spontaneously due to random fluctuations.

\end{document}